\documentstyle{elsart}
\begin{document}

\def\la{\mathrel{\mathchoice {\vcenter{\offinterlineskip\halign{\hfil
\(\displaystyle##\)\hfil\cr<\cr\sim\cr}}}
{\vcenter{\offinterlineskip\halign{\hfil\(\textstyle##\)\hfil\cr
<\cr\sim\cr}}}
{\vcenter{\offinterlineskip\halign{\hfil\(\scriptstyle##\)\hfil\cr
<\cr\sim\cr}}}
{\vcenter{\offinterlineskip\halign{\hfil\(\scriptscriptstyle##\)\hfil\cr
<\cr\sim\cr}}}}}
\def\ga{\mathrel{\mathchoice {\vcenter{\offinterlineskip\halign{\hfil
\(\displaystyle##\)\hfil\cr>\cr\sim\cr}}}
{\vcenter{\offinterlineskip\halign{\hfil\(\textstyle##\)\hfil\cr
>\cr\sim\cr}}}
{\vcenter{\offinterlineskip\halign{\hfil\(\scriptstyle##\)\hfil\cr
>\cr\sim\cr}}}
{\vcenter{\offinterlineskip\halign{\hfil\(\scriptscriptstyle##\)\hfil\cr
>\cr\sim\cr}}}}}
\def\PACS{\par\leavevmode\hbox {\it PACS:\ }}
\begin{frontmatter} 

\title{
 Spatial chaos in  weakly dispersive and viscous media: \\ 
 a nonperturbative theory of the driven KdV-Burgers equation.}
\author{M.A. Malkov\thanksref{IKI}}
\address{Max-Planck Institut f\"ur Kernphysik, D-69029, Heidelberg, 
Germany}
\PACS{47.52.+j, 52.35.Mw}
\thanks[IKI]{to appear in Physica {\bf D} \\
Email: Malkov@boris.mpi-hd.mpg.de}

\begin{abstract}
The asymptotic travelling wave solution of the KdV-Burgers equation
driven by the long scale periodic driver is constructed.  The solution
represents a shock-train in which the quasi-periodic sequence of
dispersive shocks or soliton chains is interspersed by smoothly varying
regions.  It is shown that the periodic solution which has the spatial
driver period undergoes period doublings as the governing parameter
changes.  Two types of chaotic behavior are considered.  The first type
is a weak chaos, where only a small chaotic deviation from the periodic
solution occurs.  The second type corresponds to the developed chaos
where the solution ``ignores'' the driver period and represents a
random sequence of uncorrelated shocks. In the case of weak chaos the
shock coordinate being repeatedly mapped over the driver period moves
on a chaotic attractor, while in the case of developed chaos it moves
on a repellor.

Both solutions depend on a parameter indicating the reference shock
position in the shock-train.  The structure of a one dimensional set to
which this parameter belongs is investigated.  This set contains
measure one intervals around the fixed points in the case of periodic
or weakly chaotic solutions and it becomes a fractal in the case of
strong chaos.  The capacity dimension of this set is calculated.
\end{abstract}
\end{frontmatter}
	
\section{Introduction}
Both the Korteweg de Vries (KdV) and the Burgers equations are widely
used in the study of weakly nonlinear wave phenomena in a variety of
continuous media.  These equations have the same type of nonlinearity
and represent dispersive and viscous system, respectively. In many
realistic cases, however, both viscosity and dispersion are equally
important (see eg.\cite{leib,karp,sag} and references therein) and a
generalization of these two equations known as the Korteweg-de Vries
-Burgers equation (KdVB) is very useful approximation to study the wave
dynamics.

For wave motion to appear the equilibrium state of the system must be
disturbed.  Traditionally, this is attributed to initial conditions for
the undriven equations.  As far as the KdV and the Burgers equations
are concerned, the time asymptotic behavior of the solution can often
be described in terms of the dynamical evolution of fundamental
solutions of these two equations, which are the cnoidal waves (the
solitons in the limiting case) and the shocks for the KdV and for the
Burgers equation, respectively (see e.g. \cite{whith,gur:pit}).

Different possibilities arise in the case of a continuously acting
driver, which can be of an external or a growth rate type.  The
solutions of the undriven KdV and Burgers equations can also play a
certain role when the driver is switched on.  The driven equation
itself, however, can possess a fundamental solution which is completely
different from any solution of the undriven equation.  Such a solution,
called shock-train, has been proposed \cite{ken:malk,mkwps} in the
framework of the so called derivative nonlinear Schroedinger-Burgers
(DNLSB) equation, which is the viscous version of the DNLS-equation,
introduced by Rogister, \cite{rog}.  The idea behind this solution can
be described as follows.  In the case when both the dispersion and the
viscosity are small,  the solution is defined throughout most of the
driver period by the balance between the nonlinear term and the
driver.  Since the nonlinearity of the equation provides more than one
equilibrium state (two in case of the KdVB and three in case of the
DNLSB equation, depending on the order of the nonlinear term), the
overall solution can in principle be composed of the pieces of these
different equilibria, connected by thin regions (shocks) where high
derivative terms take part in the total balance.  In terms of the
ordinary differential equation (ODE) to which the original evolution
equation is reduced when one is looking for the travelling wave
solution, these shocks correspond to the transition between the
unstable and stable types of equilibrium.  Repeating this in the
subsequent periods of the driver, we obtain a sequence of shocks, the
shock train.  Depending on the values of dispersion and viscosity each
shock can be very oscillatory and can look more like a chain of
solitons than a shock.  The major difficulty here is to define the
shock locations relative to the driver phase.  This problem can be put
in the following context.  Suppose we know the location of the shock,
if any, within a given driver period which is equivalent to the
specification of the boundary condition for the ODE.  The question then
is, where is the shock located within the neighbouring period? Provided
that we can answer this question, the solution can in principle be
calculated everywhere.

For the  Burgers equation the answer is very simple: each shock must be
at the same position relative to the driver phase, which yields a
strictly periodic solution having the driver period \cite{mss}.  This
can be understood from Poincar\'{e}-Bendixson theorem in terms of the
insufficient dimensionality of the related ODE.  Indeed, the dynamical
system that describes this travelling wave solution is two dimensional,
and according to Poincar\'{e}-Bendixson theorem its attractors are
limited by the fixed points and the limit circles.  More extensive
discussion of this issue and the relation to another type of solution
that consists of choosing a stable branch of driver-nonlinearity
equilibrium, can be found in \cite{mkk}.  The latter solution,
containing no shocks at all has been constructed in \cite{tan} in the
framework of the periodically driven KdV equation.  In the present
paper we demonstrate the possibility of a completely different response
to an external force,  giving rise to the formation of a shock-train.
Moreover, this response seems to be more persistent, due to its ability
to dissipate the pumped energy even at low viscosity, since shocks are
involved.

Like other methods of description of complicated nonlinear systems, the
shock-train technique outlined above can in principle produce
physically irrelevant solutions.  In particular, the shock train method
reduces the continuous solution to a Poincar\'e map of the shock
coordinate.  This map is no longer constrained by the
Poincar\'e-Bendixson theorem (even when the original dynamical system
is) and in principle can falsify spatially chaotic behavior.  Another
danger is the possibility of the artificial temporal chaos caused by an
insufficient number of modes in the computation of the shock train
dynamics.  For example such chaos has been observed during the
numerical study of the unstably driven Burgers equation \cite{mkk}.
This observation is interesting in view of the remarkable and, after
Lorenz`s \cite{lor} famous paper widely exploited fact, that finite
mode representation of continuous media can reveal chaotic behavior
even for a very modest number of modes.  In the case of shock-train
dynamics the number of modes must be rather large to properly describe
a shock structure at viscous and dispersive scales.  The truncation
virtually introduces the strong damping at the short scales which
facilitates the appearance of stochastic attractor and rules out the
shock-train, whereas the full solution (i.e.  calculated with
sufficiently large number of modes) shows that only regular shock-train
dynamics occurs.  Coming back to the possibility of spatial chaos, we
note that the KdVB travelling waves are governed by the driven-damped
nonlinear oscillator equation (see eq.(4) in the next section) which
has a three dimensional phase space, and hence is not constrained by
the Poincar\'e-Benidixson theorem in contrast to the Burgers equation.

A similar problem of spatial chaotization of periodically driven Alfven
waves has been studied numerically in the framework of the DNLSB by
Hada et al.  \cite{had}.  In the present paper we turn to the KdVB
equation which is a subset of the DNLSB and in its context describes
weakly nonlinear oblique MHD waves.  We also restrict ourselves to the
case of small viscosity and dispersion, which is interesting for two
reasons.  First, this case corresponds to a quite typical situation in
study of turbulence when the energy is pumped into a system at long
scales, then transferred over an inertial range to short scales via
nonlinear interaction of modes, and finally is dissipated there due to
viscosity or any other short scale dissipation mechanism.  Second, this
case is difficult for numerical integration since  different spatial
scales appear. A direct manifestation of spatial chaos in the time
dependent numerical study would formally require an infinitely large
computation box or infinite number of modes.

The plan of the paper is as follows.  In section 2 the shock-train
solution to the KdVB equation within a driver period is obtained.  In
section 3 the solution from two neighbouring periods are matched.  In
section 4 the mapping of the shock coordinate at the driver period is
introduced.  In section 5 a route to chaos through the period doubling
bifurcations is obtained as well as a developed chaos in the mapping of
the shock coordinate is considered.  Section 6 deals with numerical
integration of the driven KdVB equation and compares the results.

\section{Solution within one driver period}
\subsection{Overall behaviour}
The KdVB equation that describes nonlinear waves, driven by a source moving 
with the constant velocity can be written in the following form
	\begin{equation}
		\frac{\partial u}{\partial t} + \frac{\partial }{\partial x} 
		(u^2-\lambda^2u_{xx}-\mu u_x)=Q^{\prime} (x).
	\end{equation}
We have transformed the coordinate system to the reference frame where a 
periodic driver \[Q^{\prime} \equiv dQ/dx \] is at rest; \(u\) and \(x\) 
are normalized in such a way, that \[Q(x+2\pi)=Q(x),\] and the function 
\(Q\) is of order unity whereas the dispersion coefficient \(\lambda^2\) 
and the viscosity \(\mu\) are both assumed to be small.  We removed the 
linear term $c\frac{\partial u}{\partial x} $ from the l.h.s.  of eq.(1) by 
shifting $u \rightarrow u - c/2$.  From the physical point of view this 
choice of $x $ and $u $ variables corresponds to the case when the driver 
moves at the speed of a linear nondispersive undamped ($ \lambda ,\mu 
\rightarrow 0$) wave and, therefore, is in resonance with it.  However, 
this does not mean the loss of generality, unless $u$ is constrained by 
additional conditions like $x \rightarrow \pm \infty$ asymptotics.  Note, 
that this particular constraint is relevant rather for the unforced case or 
for the case of localized forcing.  In the case of periodic forcing, 
considered here a more suitable constraint is the mass invariant
\begin{equation}
\bar u =\frac{1}{2\pi n} \int_0^{2\pi n}u dx 
		\label{u:bar}
\end{equation}
that can be introduced for the $2\pi n$- periodic (n is integer) solution 
and is affected by the shift in $u$.  In the above context Eq.(1) remains 
generic for all steady drivers unless $\bar u $ in Eq.(\ref{u:bar})is 
specified.  Below we specify explicitely a class of particular solutions of 
Eq.(1) considered throughout the most of this paper and we shall briefly 
return to their dependence upon the mass invariant (\ref{u:bar}) in sec.6, 
where these solutions are verified by the numerical integration of Eq.(1).
      
The nondispersive shock-train solutions do not allow the spatial 
chaotization, as it was stressed in the preceding section.  We therefore 
consider the case of relatively strong dispersion
    \begin{equation} 
	\mu \ll \lambda \ll 1
	\end{equation} 
when the developed oscillatory structures must appear nearby the shocks.  
At the same time, intending to obtain a solution with distinguished shocks 
in the shock-train we assume that those oscillations are appreciably damped 
between two shocks.  The spatial damping rate of these oscillations is 
defined by $\mu /\lambda^2 $ and we assume that
	\begin{equation}
  		\beta \equiv \mu/ \lambda^2  \ga 1.
	\end{equation}
Note, that under the conditions (3,4) an efficient method for Eq.(1) would 
be that one, developed by Whitham \cite{whith:p}, and its generalization to 
the viscous driven case seems to be tractable.  However, seeking the 
travelling wave solution of Eq.(1), which moves at the driver speed, i.e.  
assuming \[\partial u/\partial t=0,\] we immediately arrive at the ODE 
which is the equation of the driven-damped nonlinear oscillator
\begin{equation}
u^2-\lambda^2u_{xx}-\mu u_x=Q(x). \label{u:sq}
\end{equation}
Our analytical treatment  in what follows will concern only (\ref{u:sq}), but 
we shall return to Eq.(1) in sec.6.

Interesting in the shock-train solutions of Eqs.(1) and (5), we choose the 
integration constant in the r.h.s.  of (\ref{u:sq}), as
\begin{equation}
Q(x)=Q_0(x)+\lambda C_1+\mu C_2+\cdots,
\label{Q:ser}
\end{equation}
where \(C_i\) are constants of order unity.
According to the general shock-train method \cite{mss,mkk} these 
constants should be chosen in such a way, that \(Q_0(x)\geq 0\) and 
\(Q_0,Q_0^\prime\) go to zero simultaneously.  This admits a pair of inviscid 
solutions \(\pm \sqrt{Q_0}\) (to the leading order in $\lambda $ and $\mu 
$) and the smooth transition between them at points 
\(x_0 \pmod {2\pi}\) where
		\[Q_0(x_0)=Q_0^\prime (x_0)=0,\]
whereas each shock in the shocktrain is supposed to be somewhere between 
two such points.

To begin, we introduce the fast variable
\begin{equation}
\xi=\frac{1}{\lambda}g_1(x)+g_2+\cdots,
\end{equation}
where $g_i(x) $ are functions to be found from the further analysis.  We 
seek the solution of (\ref{u:sq}) in the following form
\begin{equation}
u=u_0(\xi ,x)+\lambda u_1+\cdots.
\end{equation}
Then, the first two orders of approximation yield
\begin{eqnarray}
u_{0}^{2}-g_{1}^{\prime 2}\frac{\partial ^{2}u_0}{\partial \xi ^2} & =
& Q_{0} 
\label{u0:eq} \\ 
2u_{0}u_{1}-g_{1}^{\prime 2}\frac{\partial ^{2}u_1}{\partial \xi ^2} 
& = & 2g_{1}^{\prime}g_{2}^{\prime}\frac{\partial ^{2}u_0}{\partial \xi^2}+
\left(\frac{\partial u_0}{\partial \xi } \right)^{-1}\frac{\partial}
 {\partial x}g_{1}^{\prime} \left(\frac{\partial u_0}{\partial\xi}\right)^{2}
\nonumber \\ 
+ \beta g_{1}^{\prime } \frac {\partial u_0}{\partial\xi}+C_1
\label{u1:eq}
\end{eqnarray}
The solution of Eq.(\ref{u0:eq}) is a cnoidal wave (e.g.  \cite{whith}), 
which depends on \(x\) as on a parameter.  For convenience of the further 
analysis, we first represent it in the following form
\begin{equation}
u_{0}(\xi,x)=2 \sqrt{Q_0} 
\left[\zeta_{-1}+(\zeta_{0}-\zeta_{-1}) {\rm sn}^{2}\left(Q_{0}
^{\frac {1}{4}}\sqrt{\frac {(\zeta_{1}-\zeta_{-1})}{3}}\frac 
{\xi}{g_{1}^{\prime }},k \right) \right],   \label{u0:sol}
\end{equation}
where \(\zeta_n\) are the roots of the cubic equation
\begin{equation}
\zeta ^{3}-\frac {3}{4}\zeta +\frac {a}{4}=0,  \label{zeta^3}
\end{equation}
\begin{equation}
\zeta_{n}=\sin \left[\frac{1}{3}\arcsin a +\frac {2\pi n}{3} \right],
\label{zeta:n}
\end{equation}
where \(n=0,\pm 1\), and \(k\) denotes the modulus of the elliptic function 
in Eq.(11)
\begin{equation}
k^{2}=\frac{\zeta_{0}-\zeta_ {-1}}{\zeta_{1}-\zeta_{-1}}. \label{k^2}
\end{equation}
We have introduced the function \(\mid a(x) \mid \leq 1\) or, equivalently, 
\(0 < k(x) \leq 1\) which is related to \(a(x)\) by 
Eqs.(\ref{zeta^3}-\ref{k^2}), to be determined as a constant of integration 
of Eq.(\ref{u0:eq}).  In order to avoid the appearance of secular terms in 
the r.h.s.  of Eq.(\ref{u1:eq}), we specify \(g_{1}(x)\) in such a way that 
the period of the solution \(u_0 (\xi ,x) \) in \(\xi\) is independent of 
\(x\).  Then, the \(x\)-dependence of \(\partial u_0 /\partial \xi\) in the 
r.h.s.  of Eq.(\ref{u1:eq}) will not produce a nonperiodic in \(\xi\) term.  
Since
\begin{equation}
\zeta_{1}- \zeta_{-1} =\frac{3}{2} \frac{1}{\sqrt {k^2+k^{\prime 4}}},
\end{equation}
we must set in Eq.(\ref{u0:sol})
\begin{equation}
g_{1}^{\prime}=\frac{Q_0^{\frac {1}{4}}}{2^{\frac {3}{2}} {\bf 
K}(k^2+k^{\prime 4})^{\frac {1}{4}}},
		\label{g1:prime}
\end{equation}
where \(k^{\prime 2} \equiv 1-k^2\) and \({\bf K}\) is the complete elliptic 
integral of the first kind. Eq.(\ref{u0:sol}) then transforms to the following
\begin{equation}
u_0(\xi,x)=\sqrt {\frac{Q_0(x)}{k^2+k^{\prime 4}}}
\left[ -1-k^2+3k^2  {\rm sn}^2 \left(2{\bf K}(k) \xi,k) \right) \right]
		\label{u:0}
\end{equation}
To determine the function \(k(x)\) we multiply Eq.(\ref{u1:eq}) by \(\partial {u_0}/
\partial \xi\) and integrate over one period regarding \(x\) as a parameter. 
This yields the following simple equation for \(J(x)\)
\begin{equation}
\frac {\partial J}{\partial x}+\beta J=0,
		\label{j:eq}
\end{equation}
where
\begin{equation}
J(x)= g_1^{\prime } \int \limits_{0}^{1} \left (\frac {\partial 
u_0}{\partial \xi}\right)^2 d\xi.
\end{equation} 
Substituting Eqs.(\ref{g1:prime}) and (\ref{u:0}) \(J(x)\) rewrites
\begin{equation}
J(x)=\frac {6 \sqrt 2 Q_0^{5/4} (x)}{5 (k^2 +k^{\prime 4})^{1/4}} 
\left [2{\bf E} (k)-\frac{k^{\prime 2}(1+k^{\prime 2})}{k^2
+k^{\prime 4}}{\bf K} (k)\right],
	\label{j}
\end{equation}
where {\bf E} is the complete elliptic integral of the second kind.  
Eqs.(\ref{j:eq}) and (\ref{j}) implicitly define the function \(k(x)\) in 
the solution (\ref{u:0}).  Note, that the integration constants in the 
solutions of Eqs.(\ref{g1:prime}) and (\ref{j:eq}) are not specified yet.  
It is convenient to attribute one of them to the shock coordinate \(x_*\) 
so that from Eqs.(7) and (\ref{g1:prime}) we obtain
\begin{equation}
\xi \simeq \frac {g_1(x)}{\lambda}=\frac {1}{2^{3/2} \lambda} 
\int_{x_*}^{x}
\frac {Q_0^{\frac {1}{4}} dx}{(k^2+k^{\prime 4})^{\frac {1}{4}} {\bf K}(k)}.
\label{xi}
\end{equation}
The function \(k(x)\) is defined by the following equation, which results
 from Eqs.(\ref{j:eq}) and (\ref{j}) 
\begin{equation}
\frac {Q_0 ^{\frac {5}{4}} (x) Q_0^{-\frac {5}{4}}(x_* 
^{\prime})}{2(k^{\prime 2}+k^4)^{\frac {1}{4}}} \left [2{\bf E} (k)-\frac 
{k^{\prime 2}(1+k^{\prime 2})}{k^{\prime 2}+k^4} {\bf K} (k) \right]=
e^{-\beta (x-x_* ^{\prime})},
	\label{k:eq}
\end{equation}
where \(x_*^{\prime } \) represents an arbitrary constant in the
solution of Eq.(\ref{j:eq}), which we have specified in such a way that
\( k(x_*^{\prime})=1 \), whereas \( k(x_*)<1 \) (see Fig.1), and \(x_*
\) is the nearest to the point \(x_* ^{\prime}\) root of the function
\(\partial{u_0}/\partial \xi\).  The solution (\ref{u:0}) with $k(x)$
defined by Eq.(\ref{k:eq}) cannot be continued to the left from $x =
x_*^\prime $, since $k$ exceeds unity in this region and (\ref{u:0})
should be replaced there by a different solution (see Eq.(\ref{t:as})
below). However, as we will see Eq.(\ref{u:0}) yields the correct
asymptotics in the region $x_*^\prime - x \ga \lambda $, provided that
$k$ is set to unity at $x < x_*^\prime$. Therefore, we can define $k(x)$ for
$x \in (x_0 -2\pi,x_0] $
as follows
\begin{equation}
 k= \left \{ 
\begin{array}{ll} k(x), \, & x_* ^{\prime}<x \leq x_0 
\\1, & x_0 -2\pi <x \leq x_* ^{\prime},
\end{array} \right.
 	\label{k:case}
\end{equation}
\input boxedeps.tex
\SetOzTeXEPSFSpecial
\SetDefaultEPSFScale{700}
\HideDisplacementBoxes 
\begin{figure}[tbp]
\TrimBottom{14cm}
\centerline{\BoxedEPSF{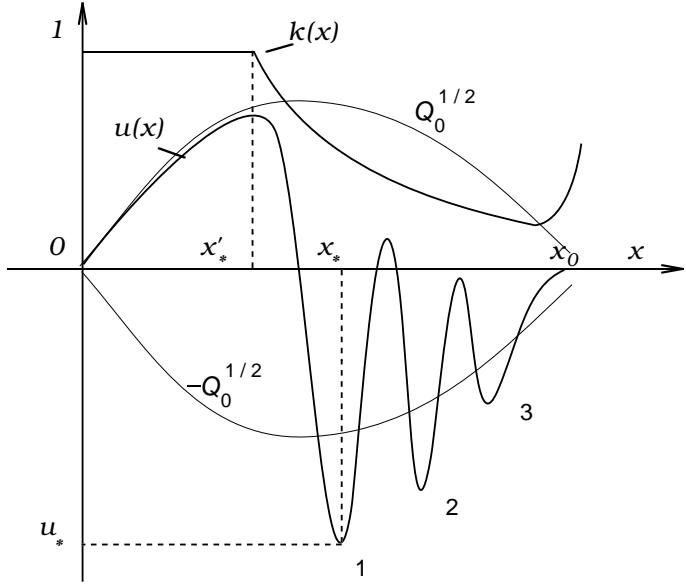}}
\caption{An arrangement of the points \(x_*\) and \(x_*^\prime \) in the 
zero order solution (17).  Three solitons marked by digits fill up 
an interval between the shock front and the `turning point' \(x_0\).  
Inviscid nondispersive solutions $\pm Q_0^{1/2} $ are plotted with thin 
lines.}
	\protect\label{fig1}
\end{figure}
where \(k(x)\) is the solution to Eq.(\ref{k:eq}) at \(x>x_* ^{\prime}\).  
Indeed, since sn in Eq.(\ref{u:0}) can be replaced in the region \(x<x_* \) 
by tanh, (by virtue of \(k^{\prime 2} (x_*) \sim \sqrt {\mu / \lambda} \ll 
1\), see Eq.(\ref{k:4}) below) the solution \(u_0\) tends to the unstable 
(for growing \(x\)) branch of the inviscid nondispersive solution of 
Eq.(\ref{u:sq}) \(u \sim \sqrt Q\).  At \(x>x_* \) it oscillates around the 
stable branch \(u \sim -\sqrt Q\), Fig.1.  Not too far from the points 
\(x_* \simeq x_*^\prime \), i.e.  at \(x - x_*^\prime < 1/\beta \), where 
\(k(x)\) is still close to unity, these oscillations virtually form a 
soliton chain attached to the shock front to the right from \(x_*^\prime 
\).  The number of these solitons is defined by the ``optical length'' 
between \(x_*\) and \(x_0\).  Beyond \(x_0\) the oscillatory behavior of 
the solution is changed for the monotonic one (see the next section).
\subsection{Resolving unsmooth region}
Although the solution (\ref{u:0},\ref{k:case}) is smooth in the fast 
variable \(\xi\), it is not smooth in the slow variable \(x \) at \(x = 
x_*^\prime \simeq x_*\).  The adiabatic approximation which allows the 
introduction of the fast and the slow variable breaks down there (the 
period of the solution in \(\xi\) tends to infinity near the separatrix
\(k=1\)), although the unsmootheness of \(u_0(x)\) does not occur in the 
main order of approximation in \(\lambda \ll 1\).  A condition for the 
smooth matching of the solution at \(x=x_*^{\prime}\), where the derivative 
of \(k(x)\) has a discontinuity, links the constants \(x_*\) and \(x_* 
^{\prime}\).  In order to do this matching we introduce the inner variable 
\(t\) in (\ref{u:sq}) defining it by
\begin{equation}
t=\frac {x-x_*}{\lambda}. \label{t:in}
\end{equation}
Eq. (4) then rewrites
\begin{equation}
u^2-u_{tt}-\frac {\mu}{\lambda}u_t =Q(x_*+\lambda t).  \label{eq:usq}
\end{equation}
Our nearest goal is to construct a part of the homoclinic (in terms of the 
slow variable $x $) orbit of Eq.  (\ref{eq:usq}) which tends to the 
solution (\ref{u0:sol}) at large \(|t| \gg 1\) for \(\mid \lambda t \mid 
\ll 1\), and which is heteroclinic in fast variable $t $ going from the 
saddle point to the stable focus (Fig.2).  Introducing the ``energy 
integral'' of Eq.(\ref{eq:usq})
\begin{eqnarray}
E  =  \frac {1}{2}u_t^2-\frac {1}{3}u^3+uQ(x_*+\lambda t)- \lambda \int
Q^{\prime}(x_*+\lambda t)u dt & + & \frac {\mu}{\lambda}\int u_t ^2 dt 
\nonumber \\ & = & {\rm const}, 
\label{energy}
\end{eqnarray}
it is useful to evaluate Eq.(\ref{eq:usq}) to the following integral for 
the function \(t(u)\).  To the first order of approximation in \(\mu 
/\lambda \ll 1\) this function can be written down as
\begin{equation}
t=-\int_{u_*}^{u} \frac {du/\sqrt 2}{\sqrt {E-uQ_*+\frac {1}{3} u^3-
\frac {\mu}{\lambda} \int_{u_*}^u u_t du}}.
\label{t:int}
\end{equation}
\begin{figure}[tbp]
\TrimBottom{14cm}
	\centerline{\BoxedEPSF{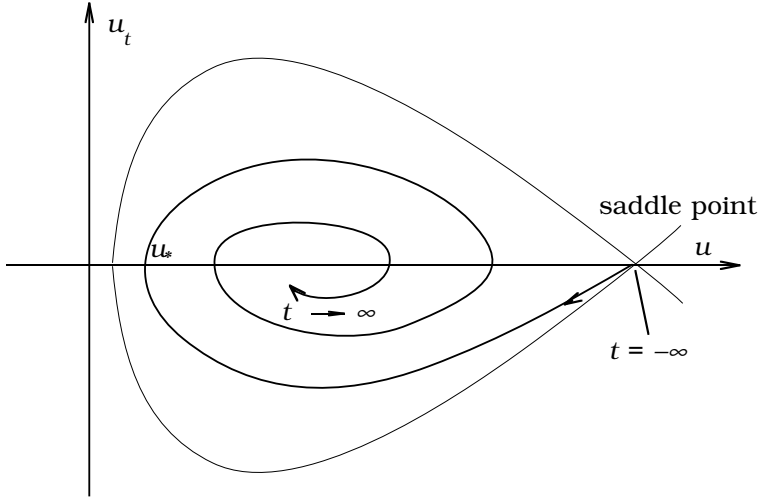}}
	\caption{Phase plane of Eq.(25)}
	\protect\label{PhPl}
\end{figure}
Here \(Q_*\equiv Q(x_*)\), \(u_* \equiv u(t=0) \) and the integral must be 
taken along the orbit with changing the branches of the square root at the 
turning points, according to the sign of $u_t $.  The choice of the 
branches along the orbit can be understood with the help of Fig.2.  We have 
omitted the terms of order of \(\lambda\) under the square root in 
(\ref{t:int}) compared to \(\mu /\lambda\) by virtue of our assumption 
\(\lambda ^2 <\mu\).  Iterating Eq.(\ref{t:int}) we substitute \(u_t\) from 
(\ref{energy}) at \(\lambda, \mu /\lambda \rightarrow 0\) which yields for 
\(u_t\) as zeroth order solution of Eq.(\ref{t:int}):
\begin{equation}
u_t^{(0)}=   \sqrt 2 \sqrt {E-Q_* u+\frac {1}{3}u^3}.
\end{equation}
Using normalized variable
\begin{equation}
\zeta = \frac {u}{2 \sqrt {Q_*}},
\end{equation}
instead of \(u\) and introducing 
\begin{equation}
S= \frac {8}{\sqrt 3}Q_*^{\frac {5}{4}}
\int_{\zeta _*}^{\zeta}d\zeta \sqrt {\zeta ^3-\frac {3}{4} \zeta +
\frac {a}{4}},
\end{equation}
from Eq.(\ref{t:int}) we have  as the second iteration (for negative \(t\))
\begin{equation}
t(\zeta)=- \frac {\sqrt 3}{2Q_*^{\frac {1}{4}}} \int_{\zeta _*}^{\zeta}
\frac {d \zeta }{\sqrt{¥P(\zeta)}}. \label{t:zeta}
\end{equation}
Here
\[\zeta _*= \frac {u_*}{2\sqrt {Q_*}}=\zeta_{-1}(k_*),\]
where \(k_*=k(x_*)\) and \(\zeta _{-1}\) were defined by 
Eqs.(\ref{zeta:n},\ref{k^2}) with
\[a= \frac {3E}{2Q_*^{\frac {3}{2}}},\]
\begin{equation}
	P(\zeta)={\zeta ^3-\frac {3}{4} \zeta +
\frac {1}{4}a+\frac {3}{8} \frac {\mu}{\lambda}Q_*^{-\frac {3}{2}}S(\zeta)}
\end{equation}
In order to be matched with the solution (\ref{u0:sol}) at \(t \rightarrow 
-\infty\), the solution \(\zeta (t)\) of Eq.(\ref{t:zeta}) must tend to the 
saddle point when \(t \rightarrow -\infty\) (see Fig.2).  This means that 
both \(P(\zeta)\) and \(dP/d\zeta\) must vanish at that point, which yields 
for the constant \(a\)
\begin{equation}
a \simeq 1-\frac {18 \sqrt2}{5Q_*^{\frac{1}{4}}}\frac {\mu}{\lambda}.
\label{a}
\end{equation}
Together with the expansions at \(k=k_*\simeq 1 \)
\begin{equation}
\zeta _* \equiv \zeta_{-1}(k_*)=-\frac {1}{2} \frac {1+k_*^2}{\sqrt 
{k_*^{\prime 2}+ k_*^4}} \simeq -1+ \frac{3}{8} k_*^{\prime 4}, 
\label{zeta_*}
\end{equation}
and
\begin{equation}
\zeta_{-1}={\rm \sin} \left(\frac{1}{3} {\rm \arcsin } (a)-\frac {2\pi}{3} 
\right) \simeq -1+ \frac{1-a}{9},
\end{equation}
it defines the constant \(x_*^{\prime}\) in Eq.(\ref{k:case}) through the 
magnitude of \(k^{\prime 2} (x) \) at \(x = x_*\), i.e through 
\(k_*^{\prime} \equiv k^{\prime}(x_*)\), which is
\begin{equation}
k_*^{\prime 4}=\frac {16 \sqrt 2}{15}Q_*^{-\frac {1}{4}} \frac 
{\mu}{\lambda}.
\label{k:4}
\end{equation}
Under this choice of \(x_*^\prime\) the asymptotics of the solutions 
(\ref{u:0}) at \(\xi \rightarrow -\infty\) and (\ref{t:zeta}) at \(t 
\rightarrow -\infty\) do coincide.  Indeed, at large negative \(t\) the 
main contribution to the integral (\ref{t:zeta}) comes from the vicinity of 
the point \(\zeta=1/2\).  Expanding \(S(\zeta)\) at this point, so that
 \begin{equation}
 	P(\zeta) = \left (\zeta -\frac{1}{2¥} \right )^2(\zeta -z),¥
 	\label{¥}
 \end{equation}¥ 
 where 
 \begin{displaymath}
 	z = -1+\frac{3}{2^{3/2}Q_*^{1/4}¥}\frac{\mu}{\lambda¥}¥¥
 \end{displaymath}¥
and substituting then \(a\) from (\ref{a}) for large negative \(t\) we 
obtain
\begin{equation}
t \sim -\frac {\sqrt 3}{Q_*^{\frac {1}{4}} \sqrt {\frac {1}{2}-z}}
\tanh ^{-1} \sqrt{\frac {\zeta-z}{\frac {1}{2}-z}}.
\label{t:as}
\end{equation}
Inverting this relation  for \(u(t)\) yields
\begin{equation}
u \sim 2\sqrt {Q_*} \left[ z+\left(\frac {1}{2}-z 
\right){\tanh }^2 \left(\sqrt 
{\frac{1}{3} \left(\frac{1}{2}-z \right)}Q_*^{\frac {1}{4}} t \right)
\right], \label{u:sim}
\end{equation}
which is equivalent to the solution (\ref{u0:sol}) or (\ref{u:0}), taken at 
\(k \simeq 1 \) and \({\bf K} \xi \) in the argument of sn (see Eq.(\ref{u:0})) 
expanded at \(x \simeq x_*\) to the same order of approximation.

Before we go to the extension of the solution to the neighbouring driver 
period, it is worthwhile to summarize the results of this section.  Given 
the shock coordinate \(x_* \) we obtained the solution (\ref{u:0}, 
\ref{k:case}), which is valid throughout one period of the driver.  Besides 
\(x_*\), the solution (\ref{u:0}) depends on an additional constant 
\(x_*^\prime\) (a point, where \(k(x)\) approaches unity and the adiabatic 
approximation breaks down), which is connected with \(x_*\) through 
Eqs.(\ref{k:eq}) and (\ref{k:4}).  This solution, however, does not allow 
the straightforward continuation to the nearby period of the driver, since 
at \(x = x_0\), where \(Q_0 =0 \) the adiabatic approximation breaks down 
again.  This region needs to be treated separately which provides the 
matching conditions for the solutions of the form (\ref{u:0}) in the two 
neighbouring driver periods.  Therefore, this matching creates a link 
between the two subsequent shock coordinates \(x_*\) and \(\hat x_*\), 
which is the task of the next section.

\section{Matching  the solutions} 
In order to extend our solution beyond one driver period, let us turn now 
to the point \(x=x_0\), where the type of equilibrium between the 
nonlinearity and the driver changes from the stable to unstable one.  Since 
\(u(x)\) tends to \(- \sqrt {Q_0}\) with increasing \(x - x_*\), we 
substitute
 \begin{equation} 
u=R+v   \label{u=r+v}
\end{equation}
  into (\ref{u:sq}), where \(v(x)\) is supposed to be small, \(R=\sqrt {Q_0}\)
 denotes the regular branch of the square root at \(x=x_0\) which is
 positive at \(x > x_0\) and is negative at \(x < x_0\) (recall, that
 \(Q_0(x) = Q_0^\prime (x_0) =0\)). Then, linearizing (\ref{u:sq}) for small
 \(v\), we obtain
\begin{equation}
2Rv-\lambda ^2v_{xx}-\mu v_x=\lambda ^2 R_{xx}+\mu R_x +Q_1+\cdots,
\end{equation}
where we denoted the constant 
\[Q_1\equiv Q-Q_0 \sim \lambda\]
 in Eq.(5). Introducing a new 
function
\begin{equation}
\phi=\exp \left[{\frac {\beta}{2}x}\right]v, \label{phi}
\end{equation}
we get the following equation
\begin{equation}
	\phi_{xx}-\left[\frac{2}{\lambda ^2}R+\frac{\mu ^2}{4\lambda ^4}
	\right] \phi = F(x), \label{phi:eq}
\end{equation}
where \(F(x)\) denotes
\begin{equation}
	F(x)\equiv-\exp \left[\frac{\beta }{2 }x \right]\left[
	\left(\frac{d}{dx}+\frac{\mu}{\lambda ^2} \right)\frac{dR}{dx}+
	\lambda ^{-2}Q_1\right].
\end{equation}
Since both \(\lambda \) and \(\mu \) are small, one can show that to the 
leading order of approximation the solution of the Eq.  (\ref{phi:eq}) is 
defined by its homogeneous part, which can be written down asymptotically 
as
\begin{equation}
	\phi(x)=\alpha \psi +\delta \chi,  \label{phi:dec}
\end{equation} 
where 
\begin{eqnarray}
\psi & = & \frac{c}{p^{1/4}} \exp \left[\frac{1}{ \lambda} S(x_1,x)\right], 
\nonumber \\ \chi & = & \frac{c}{p^{1/4}} \exp \left[-\frac{1}{ \lambda} 
S(x_1,x)\right],
\end{eqnarray}
and
\[	S(x_1,x)=\int_{x_1}^{x}\sqrt{p}dx,\]
\begin{equation} 
	p=2R+\frac {\mu ^2}{4\lambda ^2}, \quad
	p(x_1)=0. \label{t:p}
\end{equation}
Here \(x_1\) is the nearmost to \(x_0\) zero of the function \(p(x)\).  The 
branches of \(\sqrt{p}\) and constant \(c\) are chosen in such a way, that 
\({\rm Im}S>0 \, \mid c \mid =1\) on each Stokes line \(l_i\), originating 
at the point \(x_1\) and \( \arg (cp^{-1/4}) \rightarrow 0, \, x 
\rightarrow x_1, \, x\subset l_i\) (Fig.3).  Note, that we defined the 
Stokes lines \(l_i\) according to the condition \({\rm Re} S(x_1,x)=0, x 
\subset l_i \).  Given the shock coordinate \(x_*\) the asymptotic solution 
is known to the left from \(x_1\), in particular on the line \(l_1\).  A 
convenient representation of it can be derived from Eq.(\ref{u:0}) for \(k 
\ll 1\):
\begin{figure}[tbp]
\TrimBottom{14cm}
	\centerline{\BoxedEPSF{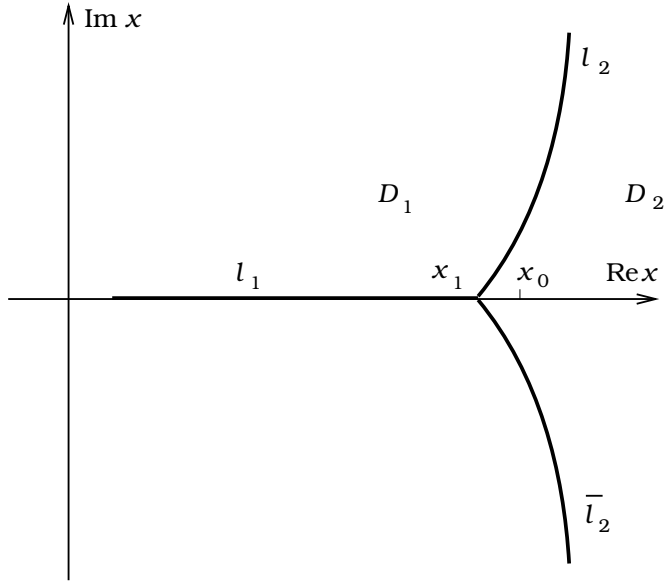}}
	\caption{The arrangement of the Stokes lines of Eq.(43)}
	\protect\label{Sl}
\end{figure}
\begin{equation}
	u_0 \simeq -\sqrt {Q_0(x)}-6\sqrt{\frac{2}{15\pi}} \frac 
	{Q_0^{5/8}(x_*^{\prime})}{Q_0^{1/8}(x)} \exp \left [-\beta (x-x_* 
	^\prime)/2 \right ]  \cos (2 \pi \xi).  \label{u0:appr}
\end{equation}
Here we have used the asymptotic solution of Eq.(\ref{k:eq}) for 
\(k^2~(x)\) in the region, where \(x\) satisfies the following two 
inequalities
\begin{equation}
\beta(x - x_*^{\prime}) \gg 1 , \, \,
x_0 - x \gg \exp\left[-\frac {2}{5}\beta (x_0 -x_*^{\prime}) \right] 
	\label{bet:ineq}
\end{equation}
The latter restriction occurs since in the vicinity of the point \(x = 
x_0\) the solution \(k^2(x)\) of Eq.(\ref{k:eq}) starts to grow becoming 
invalid because \(Q_0(x)\) approaches zero, as we noted earlier.

To make the further analysis more transparent in addition to the 
inequalities (3,4) we 
introduce the restriction \(\mu ^2 \ll \lambda ^3\), which is, however, not a 
critical one, and, at the same time corresponds to the most interesting 
case of developed oscillatory structures near the shock fronts.  Comparing 
Eq.(\ref{u0:appr}) with eqs.(\ref{u=r+v}), (\ref{phi}), and (\ref{phi:dec}), 
which represents the WKB asymptotics to the left from the turning point 
\(x_1\), we retrieve \(\alpha \) and \(\delta \) in Eq.(\ref{phi:dec})
 \[ \alpha =-2^{3/4}\sqrt {\frac 
{3}{5\pi}}Q_0^{5/8}(x_*^{\prime}) \exp \left [ \beta x_* ^\prime /2 -2 \pi 
iP_1 /\lambda \right ] \]
\begin{equation}
\delta =\bar{\alpha},
\end{equation}
	where the bar denotes complex conjugate and
\begin{equation}
P_1=\int_{x_*}^{x_1}\frac{Q_0 ^{1/4}dx}{2^{3/2} {\bf K}(x) (k^2 +k^{\prime 
4})^{1/4}}.  \label{p1}
\end{equation}
In our above calculations of the coefficients \(\alpha \) and \(\beta \) we 
have extended the asymptotic solution (\ref{u0:appr}) up to the point \(x = 
x_1\) (see Eq.(\ref{p1})).  If, however, \(x_0 - x_1 \la \exp \left 
[-\frac{2}{5} \beta (x_0 - x_*^\prime) \right ]\), which can be the case 
for relatively small \(x_0-x_*^\prime\) (note that \(x_0 - x_1 \sim \mu ^2 
/\lambda ^2 \ll 1\)), the solution (\ref{u0:appr}) together with the 
solution (\ref{u0:sol}) becomes invalid and should be replaced by the 
solution which was obtained in this section for the vicinity of the turning 
point \(x = x_1\) (eqs.(\ref{u=r+v}), (\ref{phi}), and (\ref{phi:dec})).  
According to this modification, the integrand in the phase integral 
(\ref{p1}) should be taken with \(k^2(x) = 0\) in this region.  This 
correction might become quantitatively important in calculation of the 
shock map for \(x_*\) approaching \(x_0\) (see ineq.(\ref{bet:ineq})), 
which will be taken into account in the next section.

The asymptotic solution to the right from \(x=x_1\) can be obtained by 
making use of the standard WKB approximation, which links the coefficients 
\(\alpha, \delta \) in the sectors \(D_1\) and \(D_2\) (see e.g.  
\cite{fed}).  In the region \(D_2\) the solution reads
\begin{equation}
	\phi=	\alpha _2 \psi +\delta _2 \chi 
\end{equation}
	with
\begin{equation}
		\alpha _2=e^{-i\pi /3}(\alpha+i\bar {\alpha})
\end{equation}
	and
\begin{equation}
		\delta _2=ie^{-i\pi /3} \alpha.
\end{equation}
Thus, a solution, which is valid also for real \(x>x_1\) can be 
written in terms of the function \(v(x)\) (Eq.(\ref{u=r+v})) as
\begin{equation}
v(x) = 2^{-1/4}R^{-1/4}(x)e^{-\beta x/2-i\pi/4}\left [(\alpha +i\bar 
\alpha) e^{S(x_1,x)/\lambda}+i\alpha e^{-S(x_1,x)/\lambda}\right ].
\end{equation} 
Taking \(\mu ^2\ll \lambda ^3\) into account, after a simple algebra from 
the last equation we obtain
\begin{eqnarray}
      u(x) & = & R - \sqrt{\frac {6}{5\pi}}\frac {Q_0 ^{\frac {5}{8}} (x_* 
   ^{\prime})}{Q_0^{\frac {1}{8}}(x)}e^{-\beta (x-x_* ^{\prime})/2} \cdot
   \nonumber \\ 
  & \cdot & \left [ 
   2 \cos \left ( \frac {2 \pi}{\lambda} P_1+ \frac{\pi}{4}\right ) \exp 
   \left ( \frac {\sqrt 2}{\lambda} \int_{x_1} ^x Q_0 ^{\frac {1}{4}} dx 
   \right ) \right.\nonumber \\ 
   & + & 
  \left. \exp \left (\frac {i\pi}{4}-\frac {2i 
   \pi}{\lambda}P_1-\frac {\sqrt2}{\lambda} \int_{x_1} ^x Q_0 ^{\frac 
   {1}{4}} dx\right ) \right ].
   \label{u:wkb}
\end{eqnarray}
Now we can construct the solution in the region \(x_0<x<x_0+2\pi\).  
Suppose the next shock is located at \(x=\hat x_* + 2 \pi \), so that we 
take all the subsequent shock coordinates \( x \pmod {2 \pi} \).  Our task 
is to link \(\hat{x_*}\) with \(x_*\), which defines by an induction the 
solution for all \(x>x_*\).  According to Sec.2, the solution, which is 
valid in the region \(x_0<x<x_0+2\pi \), but not too close to the end 
points of this interval can be written in the form (\ref{u:0}), with \(\hat 
x_*\) instead of \(x_*\) in (\ref{xi}).  Then, taking (\ref{u0:sol}) at 
large negative \(\xi\), i.e.  using Eq.(\ref{u:sim}) with \(\hat x_*\) 
instead of \(x_*\), we obtain
\begin{equation}
u(x) \simeq \sqrt{\hat Q_*}\left\{1-12 \exp \left[\frac {2}{\sqrt 3} 
\sqrt{\frac{1}{2}-\hat \zeta _*}\hat Q_* ^{\frac {1}{4}} \frac {x-\hat x_* 
- 2 \pi }{\lambda}\right]\right\}.
\end{equation}
Here we introduced the notations \(\hat Q_*\equiv Q(\hat x_*)\) and \(\hat 
\zeta _* \equiv \zeta _*(\hat x_*)\), whereas the function \(\zeta 
_*(x_*)\) is defined by Eq.(\ref{zeta_*}).  Expanding \(\hat \zeta _*\) at 
small \(\mu /\lambda\) yields \begin{equation} u \simeq \sqrt {\hat 
Q_*}-12\sqrt {\hat Q_*}\exp \left [\sqrt 2 \hat Q_*^{1/4}\frac {x-\hat x_* 
-2\pi}{\lambda}- \beta (x- \hat x_* - 2 \pi )/2 \right ].  \label{u(x)} 
\end{equation} 
This asymptotic solution should be matched with the 
asymptotics (\ref{u:wkb}) in the region where they both are valid.  This 
region is defined by the following two inequalities \[\lambda \ll \hat x_* 
+ 2 \pi -x \ll \hat x_* + 2 \pi - x_0,\] To match Eqs.(\ref{u:wkb}) and 
(\ref{u(x)}) we can evaluate the integral in Eq.(\ref{u:wkb}) as follows 
\[\int_{x_1}^{x}Q_0^{1/4}dx \simeq \int_{x_1}^{\hat x_* + 2 \pi 
}Q_0^{1/4}dx-\hat Q_*^{1/4}(\hat x_* + 2 \pi -x).\] The asymptotic 
solutions (\ref{u:wkb}) and (\ref{u(x)}) will coincide in the region of 
their validity if the following condition is met
\begin{eqnarray}
\frac{Q_0 ^{5/8}(x_* ^\prime )}{\sqrt {30\pi} \hat Q_* ^{5/8}} \cos 
\left(\frac {2 \pi}{\lambda } P_1+\frac {\pi}{4}\right) \cdot \nonumber \\ 
\cdot \exp \left [-\frac {\beta }{2}(\hat x_* + 2 \pi -x_* ^{\prime})+ 
\frac { \sqrt 2}{\lambda }\int_{x_1}^{\hat x_* +2 \pi } Q_0^{1/4}dx\right 
]=1.  \label{map:orig}
\end{eqnarray}
\section{Mapping the shock coordinate}
\subsection{Arbitrary \(2\pi \)- periodic driver}
Equation (\ref{map:orig}) provides an implicit relation between the two 
subsequent shock coordinates \(x_*\) and \(\hat x_*\).  However, it is 
still rather complex for a comprehensive analysis.  First of all there are 
two intermediate coordinates \(x_*^\prime \) and \(x_1\) in 
Eq.(\ref{map:orig}) that appeared during the evaluation of the map 
\(x_*\mapsto \hat x_*\).  They have no other physical meaning than \(x_*\) 
and \(x_0\) already do.  Given the shock coordinate \(x_*\), the value 
\(x_*^\prime\) can in principle be calculated by means of Eqs.(\ref{k:eq}) 
and (\ref{k:4}), whereas \(x_1\) is defined by Eq.(\ref{t:p}) and lies 
close to the point \(x_0 \).  Moreover, our assumption \(\mu^2 \ll 
\lambda ^3\) allows us to replace \(x_* ^{\prime} \) by \( x_*\) and \(x_1 
\) by \( x_0\), respectively (see Eqs.(\ref{k:eq}), (\ref{k:4}) and 
(\ref{t:p})).  We can also set \(x_0= 2 \pi \) without loss of generality.  
After these simplifications we rewrite the map \(x_*\mapsto \hat x_*\) as
\begin{equation}
		f(x_*)=g(\hat x_*), \label{f=g}¥
\end{equation}
	where
\begin{equation}
		f(x)=Q_0^{5/8} (x) \cos \left[\frac { \pi }{\sqrt 2 \lambda }
		\int_{x}^{2 \pi}\frac {Q_0^{1/4}dx}{{\bf K} (x)(k^2+k^{\prime 
	4})^{1/4}}+\frac {\pi}{4}\right ] e^{\beta (x/2-\pi)}, \nonumber
\end{equation}
and
\begin{equation}
	g(x)=\sqrt {30\pi} Q_0^{5/8}(x) \exp \left[-\frac {\sqrt 
	2}{\lambda}\int_{0}^{x}Q_0^{1/4}dx+\frac {\beta x}{2} \right].
\end{equation}
\subsection{Sinusoidal driver}
Let us consider the simplest case of sinusoidal driver in Eq.(1):
\begin{equation}
Q{^\prime} = \frac{1}{2} \sin (x) \label{Q'}
\end{equation}
Then, according to Eq.(\ref{Q:ser}) and the text right after it we have for 
\(Q_0\) \[Q_0 = \sin^2 \frac{x}{2},\]
The functions \(f\) and \(g\) are shown in Fig.4a.  One can see, that given 
\(x_*\), Eq.(~\ref{f=g}) can not be uniquely resolved for \(\hat x_*\), since 
there are two different \( x \) for each \(g(x)\).  At the same time, the 
function \(g(x)\) at \(x<x_m\) (the point of maximum of function \(g(x)\)) 
corresponds to the position of the next shock \(\hat x_*\), which is too 
close to \(x_0\) to be correctly described by our approximation, since the 
asymptotic method of the derivation of the map \(x_*\mapsto \hat x_*\) 
implies, that \(\hat x_* \gg \lambda\).  Indeed, \(x_m \sim 
\lambda ^{2/3}\) and the solution with \(\hat x_* < x_m\) can hardly 
satisfy the 
inequality \(\hat x_* \gg \lambda\) except for very small  $\lambda $.  
In general, this solution would correspond to the shockless one, which stays 
at the stable equilibrium \(u \sim -\sqrt Q\) after the phase of the driver 
advanced beyond the point \(x_0\) (the jump would be of the order of \(x_m - 
x_0 \ll 1\)) and the matching should be treated on different basis.  The 
above nonuniqueness is important when considering Eq.(\ref{u:sq}) per se.  
Here, we are looking for the travelling wave solutions of Eq.(1) and any 
particular choice of solution is justified.  Interested in the shock-train 
solutions we choose \(\hat x_* \in [x_m,2 \pi] \) for \( g(\hat x_*) 
\) from Eq.(\ref{f=g}).  In this case the mapping \(x_*\mapsto \hat x_*\) 
is unique, although it is not defined for \(x\), where
\begin{equation} 
f(x)< \min_{x\in [x_m,2 \pi]} g(x) \label{f<g}
\end{equation} 
as well as in the interval \([0,x_m)\).  Resolving Eq.(\ref{f=g}) 
numerically and omitting stars at \(x\) we get the mapping \(F(x):x \mapsto 
\hat x\), which is shown in Fig.4b.
\begin{figure}[tbp]
\TrimBottom{0cm}
	\centerline{\BoxedEPSF{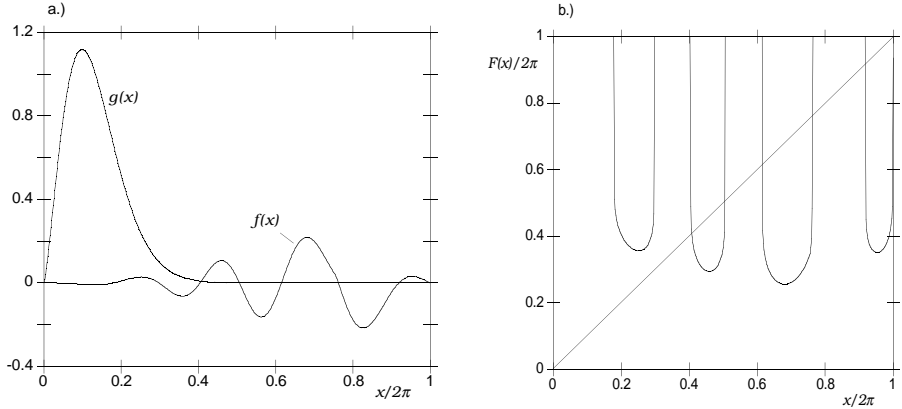}}
	\caption{a.) The functions \(f(x)\) and \(g(x)\) 
	         b.) The map of the shock coordinate \(F(x)\) }
	\protect\label{f-g}
\end{figure}
\section{Simple solutions of period one and the period-doubling bifurcations}  
\subsection{Weak chaos}
\setcounter{footnote}{0}
As discussed in the preceding section, the map\footnote{In what follows we 
relate \(x, F(x) \) to the interval \([0,1]\) instead of \([0,2\pi] \).} 
\(F(x)\) (Figs.4) is not defined in those intervals where the inequality 
(\ref{f<g}) is met.  This means that after an iterate of the map \(F(x)\) 
has fallen into one of these intervals (denote them \(J_{\rm k}\), and 
include also \([0,x_{\rm m})\) in this category, as we have mentioned 
above), the travelling wave solution of the shock-train type cannot be 
continued to the right from this point.  At the same time, a continuous to 
the zeroth order (in \(\lambda\) and \(\mu\)) solution could in principle 
be constructed in this driver period, as it was described in sect.1.  
However, we leave aside this possibility and restrict our consideration to 
the shock-trains.  According to Eq.(\ref{f=g}), the total number of 
intervals where \(F(x)\) is defined, can be estimatd as \(N_{\rm I} \sim 
1/\lambda \gg 1\).  We denote these intervals as \(\left \{ I_ k \right 
\}_{k=1}^{N_{\rm I}} \).  The number of the fixed points of the map 
\(F(x)\) is \(N_{\rm F} \la 2N_{\rm I}\).  Each fixed point of \(F(x)\) 
corresponds to \(2\pi\) -periodic solution of Eqs.(1) and (4).  Since at a 
fixed point \(x\)
\begin{equation}
\mid dF/dx \mid = \frac{f^ \prime (x)}{g^ \prime (x)} \sim \exp \left
[\frac {\sqrt{2}}{\lambda} \int_0^x Q_0^{1/4} dx \right ] \gg 1,
	\label{unst:map}
\end{equation}
these fixed points are mostly unstable with a possible exception for a 
point, where \(g(x)\) is not very small, or where \(f(x)\) reaches its 
maximum.  Accordingly, the one parameter ($x_* $) family of solutions of 
Eq.(\ref{u:sq}), constructed in the previous section with \(x_*\) taken to 
be equal to one of this fixed points represents an unstable periodic orbit 
of Eq.(\ref{u:sq}).  It is important to note here, that the unstable in 
\(x\) periodic solution of Eq.(\ref{u:sq}) can correspond to the stable (in 
time) periodic travelling wave solution of the driven KdVB Eq.(1).  This 
will be demonstrated in sec.6.  We also note, that the instability of a 
fixed point evolves in the positive \(x\)-direction, whereas the same fixed 
point attracts the backward iterations of the Poincar\'e section even 
though the map has no unique inversion.  The question how the spatial 
stability of our periodical solition is related to its temporal stability 
deserves a special treatment which is beyond the scope of this paper.  
Note, that even in linear systems the answer can be not a straightforward 
one (e.g.  absolute versus convective instability in plasmas).
\begin{figure}[tbp]
\TrimBottom{0cm}
	\centerline{\BoxedEPSF{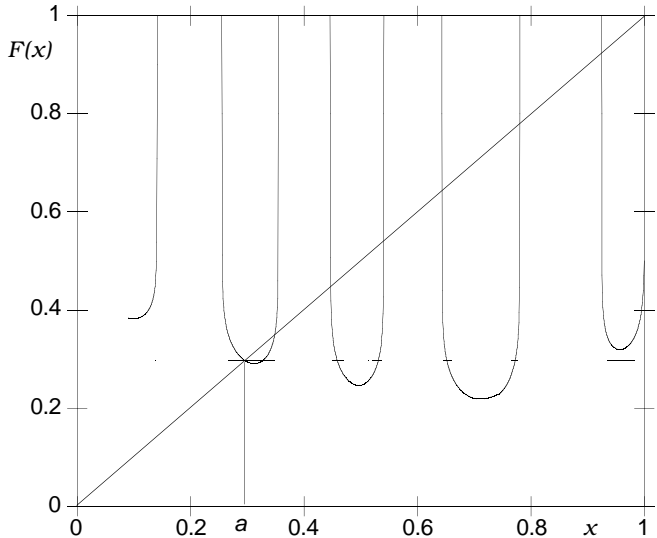}}
	\caption{Mapping of the shock coordinate with one stable fixed point \(x 
	= a \) for \(\lambda =0.27 \) and \(\beta = 1.2 \).  Horisontally aligned 
	points, which are generated by \(F^{100}(x) \),  
	actually mark the basin of attraction of this fixed point}
	\protect\label{stable}
\end{figure}

We begin our analysis of the map \(F(x)\) with the case when a minimum of 
\(F(x)\) is close to one of its fixed points, say \(x = a\), (Fig.5).  In 
this case the local behavior of the iterates near this fixed point does not 
differ from that which is known for one dimensional map with one maximum 
\cite{feig}.  In particular, starting with the stable fixed point \(x =a, 
\, \mid F^\prime(a) \mid <1 \), we obtain a series of period doubling 
bifurcations when, for instance, \(\lambda \) is fixed \(\lambda = 0.27\) 
while \(\beta\) changes from \(\beta = 1.2 \) down to \(\beta = 0.98\) (see 
Fig.5-7).  Below the point \(\beta = 0.98\) the local behavior of the 
mapping \(F(x)\) becomes chaotic.  At the tangential bifurcation of the 
fixed point one can observe an intermittent behavior of the mapping 
\cite{pom}, although the return time to the vicinity of the tangent point 
(where \(F(x) -x \ll x\)) might be quite large, since the orbit visits many 
other intervals of the mapping \(F(x)\) before it finds the return path, if 
ever.  The measure one basin of attraction of the fixed point \(x =a\) and 
emerged \(2^n \) cycles is made up of the points which are under the 
horizontal lines seen in Figs.5-6 and under the scattered points shown in 
Figs.7.  When we start to iterate \(F(x)\), with an initial point from the 
basin of attraction the iterations will be attracted by the fixed point or 
will be trapped in \(2^n\) cycle which is the fixed point of the map 
\(F^n\) defined by \(F^n = F\circ F^{n-1}(x) \); \(F^0(x) \equiv x\).  
Almost all other points of the \([0,1]\) interval will sooner or later 
escape in one of the intervals \(J_k \) where \(F(x)\) is not defined.  
This means that for almost all initial shock coordinates which are not from 
the basin of attraction, the shock-train solution can not be continued 
infinitely into the positive \(x\)-direction.  The rest (a measure zero 
subset) of the \([0,1]\) interval, which we meant by saying ``almost all'' 
will be considered in the next section.  It is worth noting, that although 
the measure zero solutions are exceptional for the ODE (\ref{u:sq}), they 
can be very persistent and quite ``typical'' travelling wave solutions of 
the correspondent evolution equation Eq.(1).
\begin{figure}[tbp]
\TrimBottom{0cm}
	\centerline{\BoxedEPSF{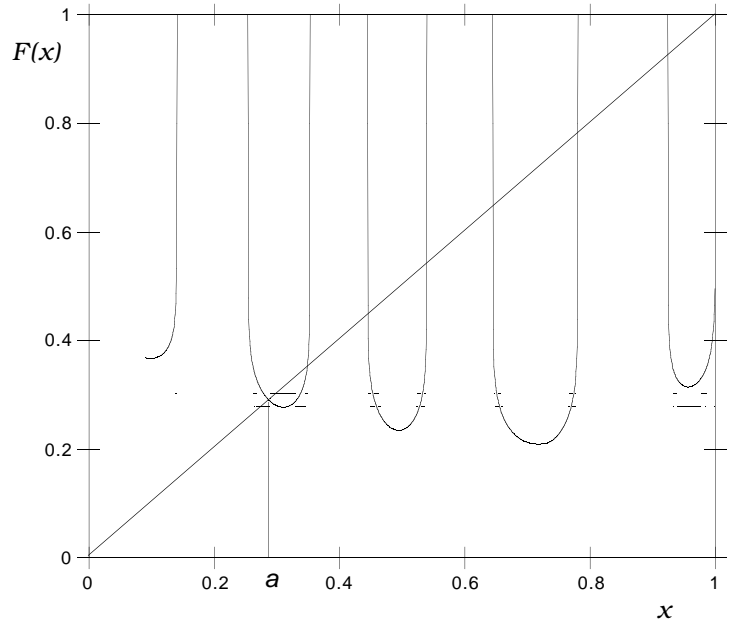}}
	\caption{Stable cycle of period 2 near the fixed point \(a\), \(\lambda = 
	0.27\), \(\beta =1.1\). The 100-th 
	(\(F^{100}(x)\)) iteration is shown as well as \(F(x)\)}
	\protect\label{DP}
\end{figure}
\begin{figure}[tbp]
\TrimBottom{0cm}
	\centerline{\BoxedEPSF{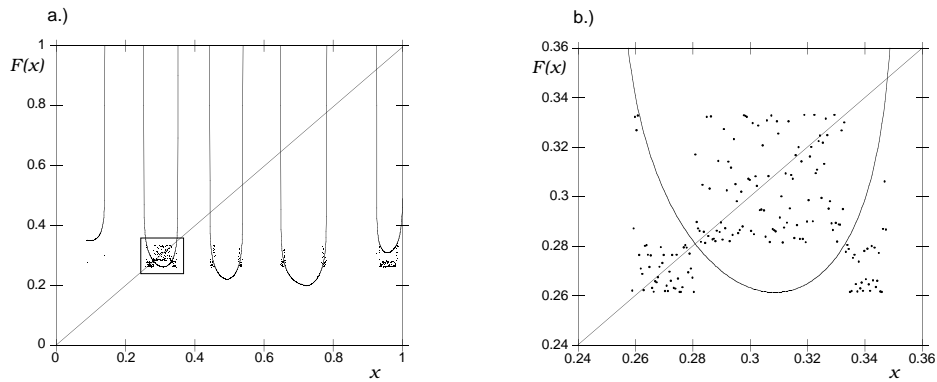}}
	\caption{a.) The same as in Fig.6 but for \(\beta = 0.98 \)  b.) A 
	magnification of the box in Fig.7a. }
	\protect\label{weakc}
\end{figure}
\subsection{Strong spatial chaos}
\begin{figure}[tbp]
\TrimBottom{0cm}
\TrimTop{3cm}
	\centerline{\BoxedEPSF{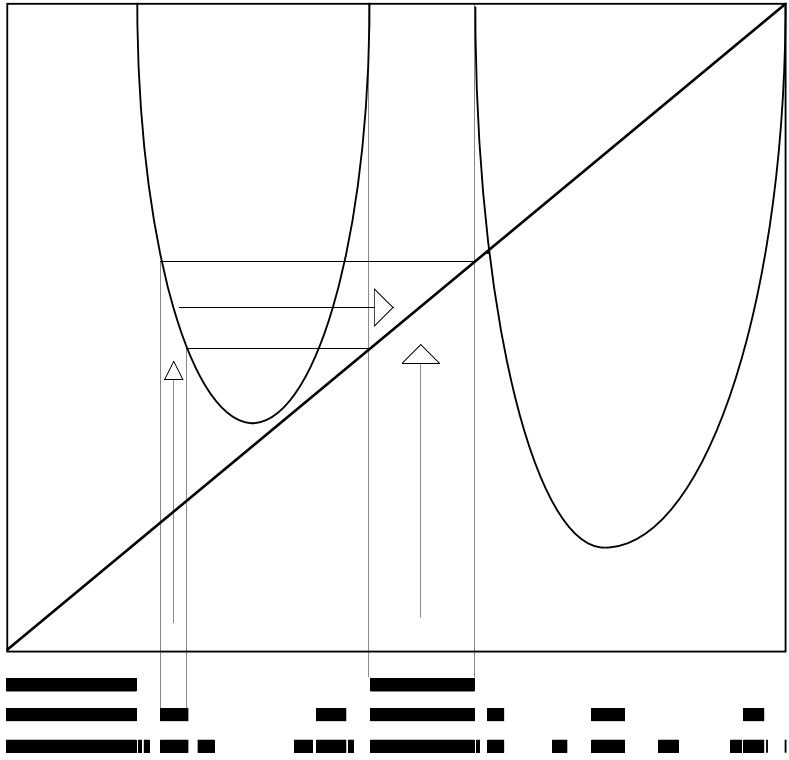}}
	\caption{A piece of the map \(F(x)\) illustrating the fragmentation of 
	phase space under iterations. The points 
	from the black areas will escape as it is shown by the arrows.}
	\protect\label{frac}
\end{figure}

The behavior of the iterates considered in the previous section is chaotic 
for \(\beta \) below \(\beta _{\rm c} = 0.98\), but this chaos is weak in 
the sense that the spatial profile of the solution of Eq.(1), corresponding 
to this sequence of shock coordinates \(\left \{x_{n} \right 
\}_{n=0}^\infty\) is still rather close to periodic solution since \(\mid 
x_{n} - a \mid \la \lambda\).  By further decreasing of \(\beta \), the 
measure one basin of this weakly chaotic attractor around \(x =a \) 
disappears, and almost all initial points will escape in 
\([0,1]\setminus\bigcup_{k=1}^{N_{\rm I}}I_k\) as it was argued in the 
previous subsection.  To identify the nonwandering points of \(F\) which 
define now the sequence of shocks in the spatially chaotic solution of 
Eq.(1), we proceed as follows.  Consider the nonescape set, which occurs 
under the backward iterations of \(\bigcup_{k=1}^{N_{\rm I}}I_k\) and 
denote its limit by \( A = [0,1] \setminus \bigcup _{n=0}^\infty F^{-n} 
\left ( \bigcup J_k \right )\), (recall, that $J_k $ denotes the escape 
intervals).  The construction of this set is similar to that of the 
classical Cantor set and is illustrated in Fig.\ref{frac}.  Once the 
initial shock coordinate is taken from the set \(A\), the shock train 
solution can be continued ad infinitum.

\begin{figure}[tbp]
\TrimBottom{0cm}
	\centerline{\BoxedEPSF{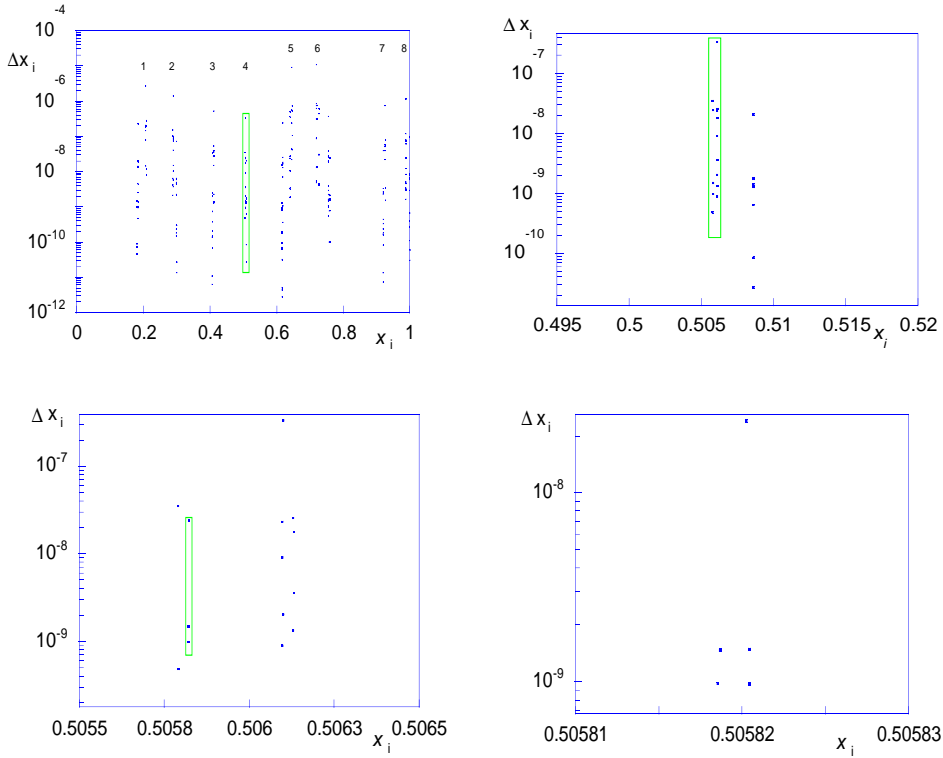}}
\caption{Illustration of the natural measure of chaotic repellor of the map 
\(F(x) \). All four plots show the lengths of intervals that have not 
escaped after six iterates of \(F\), as a function of 
their left boundary.  Each box is magnified in the subsequent 
plot.}
	\protect\label{iter}
\end{figure}

\begin{figure}[tbp]
\TrimBottom{0cm}
	\centerline{\BoxedEPSF{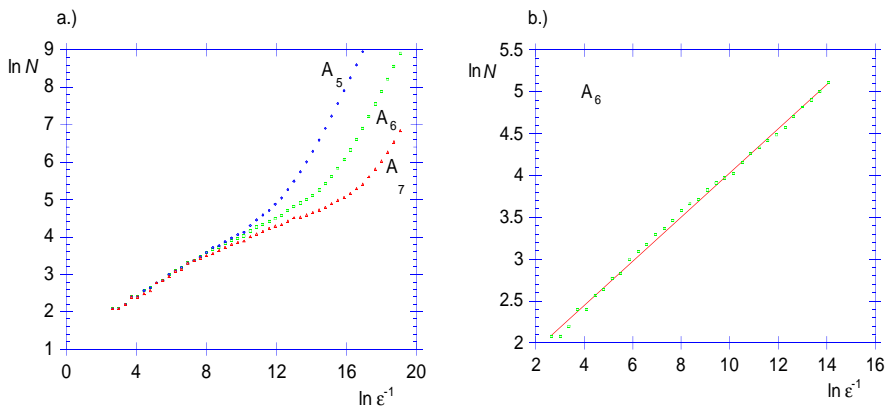}}
	\caption{a.) The number of intervals needed to cover the sets \(A_n\) for 
	\(n=5,6,7\) and 
	for \(\lambda=0.3 \), \(\beta=1.3\)
	b.) A linear fit to $\ln N $ for the case $n = 6 $.}
	\protect\label{cap:dim}
\end{figure}

An important property of this set is its capacity dimension.  One 
calculates it from the following formula, e.g.  \cite{farm,gras}
\begin{equation}
D_{\rm c} = \lim_{\epsilon \rightarrow 0} \frac{ \log N(\epsilon)}{\log 
(1/\epsilon)},
\label{dim}
\end{equation}
where \(N\) is the minimal number of the intervals of length \(\epsilon \), 
needed to cover the set.  To calculate \(D_{\rm c}\) we approximate the set 
\(A\) by iterating the interval \([0,1]\) and by removing all those \(x\), 
which after \(n\) iterations leave the domain of \(F\), i.e.  escape in one 
of the intervals \(J_k\).  Denote the rest of $[0,1] $ by \(A_n \equiv 
[0,1]\setminus \bigcup_{m=0}^{n} F^{-m} \left ( \bigcup J_k \right ) = 
\left \{ (x_i,x_i+\Delta x_i) \right \}_{i=1}^{N_s}\), so that $A = \lim_{n 
\rightarrow \infty} A_n$.  The distribution of segment lengths $\Delta x_i 
$ ($N_s $ is their total number) of the set remained after six iterations 
\(A_6\) is shown in Fig.\ref{iter} as a function of the left boundary of 
each segment.  The iterations \(A_n \) cannot be continued beyond certain 
\(n = n_{\rm max}\), because of the limited computation accuracy.  Since 
the map \(F(x)\) expands phase space very quickly (see 
Eq.(\ref{unst:map})), errors in the calculation of the points from 
invariant set of \(F\) will rapidly grow and after a few iterations the 
point escapes into one of the intervals \(J_k\) just due to insufficient 
accuracy.
Therefore, for numerical purposes we consider the sets \(A_n\) with \(n = 
5,6,7\).  After \(n \ga 8\) iterates the number of segments $N_s$ in 
\(A_n\) starts to decrease for above reasons.  The dependence 
\(N(\epsilon)\) for different \(A_n\) is shown in Fig.\ref{cap:dim}.  Since 
\(A_n\) is measure one for any finite \(n\) , even though very irregular, 
the dependence \(N(\epsilon)\) becomes \( N \sim \epsilon ^{-1}\) for 
\(\epsilon \) smaller than typical segment length in \(A_n\).  At larger 
\(\epsilon\) one can see a distinct region of the linear behavior in the 
\(\log-\log\) scale, which captures the main fractal properties of the 
limiting \((n \rightarrow \infty)\) set \(A\).  In this \(\epsilon \)- 
range the slopes of the curves are essentially identical for \(4 < n < 8\).  
The transition to the \(D_{\rm c} = 1\) behavior at smaller \(\epsilon \) 
occurs earlier for smaller \(n\) which is a natural result.  For \(n \ge 
8\) the slope becomes smaller, which underestimats the capacity dimension 
of set \(A\) due to the escape of points from \(A_n\) because of the 
limited computation accuracy.  For \(\lambda=0.3 \) and \(\beta=1.3\) the 
capacity dimension \(D_{\rm c} \approx 0.26\).
\section{Comparison with the time dependent numerical solutions}
Before we verify numerically some of our above results it is worthwhile to 
comment on the possible approaches to this verification.  First of all 
there are two different aspects of our analytic calculations.  One of them 
concerns the response of damped nonlinear oscillator (\ref{u:sq}) to the 
large amplitude forcing.  In the present paper we leave aside the problem 
of numerical study of (\ref{u:sq}) regarding its reduction to the discrete 
map (\ref{f=g}).  We only note here that such a study would not be 
straightforward, since the typical fixed point of this map is unstable as 
well as the motion on the repellor, considered in the previous section and 
the solutions of (\ref{u:sq}) are unbounded in their vast majority.  
Backward iterations of this map, locally attracted by this fixed points, 
would require a choice of preimages, i.e.  the branches of \(F^{-1}(x) \), 
since \(F(x) \) is not invertible.  Even though this numerical study were 
successively done, it would not give us an ultimate answer to an important 
question concerning the driven KdVB equation.  In particular, it would be 
still unclear which solutions of the travelling wave type, i.e.  the 
solutions of (\ref{u:sq}) are the attractors of Eq.(1)? We therefore 
perform direct integration of the evolution equation (1) in time.  It 
should be realized, however, that this approach is also insufficient for a 
firm manifestation of spatial chaos, since it would require an infinite 
number of modes or equivalently an infinitely long system.  An additional 
problem is that we already have two essentially different scales 
(dispersion/viscosity and the driver scales) and in order to detect at 
least a possible trend to the spatially chaotic behavior we would need to 
model a system that appreciably exceeds the driver scale.  Note, that it is 
the shock-train technique, developed in the previous sections that is 
especially suitable for handling systems with considerably different 
spatial scales.  At the same time, the spatially chaotic travelling wave 
solutions, obtained in the preceeding sections are not necesseraly the 
attractors of time dependent Eq.(1).  An example of the impossibility of 
chaotic solutions is given in \cite{aub} where the Frenkel-Kontorova model 
is considered.  In particular, it was proven in \cite{aub}, that the 
ground-states of this model can be periodic or quasi-periodic but the 
spatial chaos is ruled out.
  
The numerical investigation of the onset of spatial chaos is beyond the 
scope of this paper.  The above arguments suggest that it could be a rather 
difficult task and certainly deserves a separate study.  First of all one 
should check whether at least the first steps in a route to spatial chaos 
can be described correctly by an analytical shock-train solution obtained 
in previous sections.  A possible strategy consists of obtaining $2\pi $ 
periodic solutions numericallly and of further identification them with the 
correspondent fixed points of the map $F(x) $ (sec.4).  Then, the $4\pi 
$-periodic solutions should be found and identified with the fixed points 
of $F^2 \equiv F \circ F$.  These fixed points appear in the vicinities of 
the unstable fixed points of $F$.  A continuation of this process should in 
principle give rise to the spatially chaotic solutions.  This section deals 
with the very begining of this program.  First, we demonstrate that the 
travelling wave solutions of (\ref{u:sq}) do attract time dependent 
solutions of Eq.(1).  Second, an approximate Poincar\'e map can really 
capture their complexity.

We start with the numerical integration of Eq.(1) for periodic driver given 
by Eq.(\ref{Q'}) with different initial conditions.  The results are shown 
in Figs.11.  One can see from these two figures that the shock-train is a 
persistent solution of Eq.(1).  Starting from large amplitude sinusoidal 
profile (Fig.11a), the solution steepens very quickly and forms a 
shock-train which has the driver period.  Fig.11b demonstrates the case 
with completely different initial condition, where a rather random small 
amplitude perturbation is imposed.  It decays rapidly due to the dispersive 
spreading and viscosity.  Then after the amplitude of the driver's harmonic 
achieved nonlinear level, it steepens to form exactly the same 
travelling wave solution.
\begin{figure}[tbp]
\TrimBottom{5cm}
	\centerline{\BoxedEPSF{fig11.eps}}
	\caption{Formation of the shock-train for two different initial 
	conditions, and the driver given by Eq.(63) The coordinate is
	shifted back by \(\pi/2 \) and compressed to \([0,1] \). Only one period 
	of the solution is shown.}
	\protect\label{time:dep}
\end{figure}
Figs.12 show the comparison of such a steady state with an asymptotic 
solution given by the formulae (\ref{u:0},\ref{xi}-\ref{k:case}).  It is 
seen from Fig.12a that the agreement is very good, taking into account that 
only zero order solution \(u_0(\xi ,x) \) is plotted.  There are, however, 
two regions where the deviation is significant.  They correspond to the 
nonuniformities of the asymptotic solution (\ref{u:0}), resolved in secs.2 
and 3, but not taken into account in Figs.12 for simplicity.  First, the 
amplitude of the first peak is somewhat larger than numerical one due to 
our replacement \(x_*^\prime \) by \(x_* \), which overestimates \(k(x)\) 
at \(x \simeq x_* \) (see sec.2).  Second, the deviation in the region 
where an oscillatory behavior is changed for a monotonic one results simply 
from invalidity of \(u_0(\xi,x)\) given by formula (\ref{u:0}) at the 
turning point \(x_0\) (sect.3) in the case of moderate \(\beta \).  (We 
have taken \(\beta = 1.8 \) to obtain pronounced solitons between the shock 
and the turning point \(x_0\)).  Thus, the agreement can be even improved 
by using the asymptotics given in sec.2,3 instead of Eq.(\ref{u:0}).
\begin{figure}[tbp]
\TrimBottom{0cm}
	\centerline{\BoxedEPSF{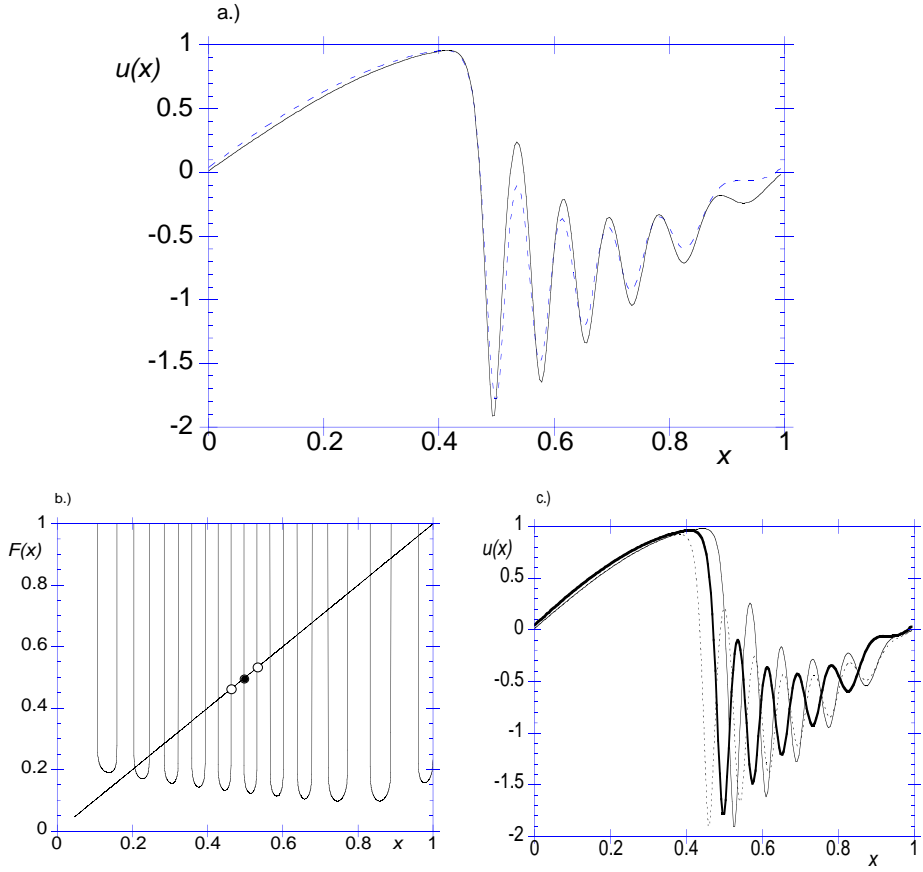}}
\caption{a.) Zeroth order analytic solution Eq.(17) (solid line) 
having the shock at \(x_*=0.497\) (the correspondent fixed point of the map 
$F(x) $ is marked by the filled circle in Fig.12b).  Dashed line shows the 
steady state numerical solution.  Both curves are drawn for the case 
\(\lambda = 0.1\) and \(\beta = 1.8\).
\newline
b.) The map \(F(x) \) for the same \(\lambda \) and \(\beta \).  c.) The 
same numerical solution as in Fig.12a (heavy line) and the two analytical 
solutions (thin and dotted lines) having shocks at \(x_*\) marked by the 
opened circles in Fig.12b.}
	\protect\label{}
\end{figure}

The solution \(u_0(\xi, x) \) depends also on additional parameter, the 
shock coordinate \(x_*\).  For the period-one solutions considered here, 
\(x_*\) must be equal to one of the fixed points of the map \(F(x) \), 
plotted in Fig.12b.  In the analytical solution shown in Fig.12a, \(x_* = 
0.497 \) which corresponds to the filled circle in Fig.12b.  Solutions 
corresponding to the empty circles in Fig.12b are shown in Fig.12c and they 
clearly do not agree with the numerical solution.  This allows us to 
identify a continuous solution with one of the discrete shock coordinates.  
There are as many as twenty fixed points of the map \(F(x) \) shown in 
Fig.12b.  Each of them corresponds to a period-one solution of the driven 
KdVB equation.  Moreover, one can easily draw \(F^n(x)\) for \(n > 1 \) and 
obtain, in principle, all the solutions of period \(n\), as the fixed 
points of \(F^n\).  Due to the limitations of applied asymptotic method, a 
few right most fixed points cannot be described correctly.  At the same 
time they correspond to the solution, in which \(u \sim \sqrt Q \) 
throughout the driver period, i.e.  actually to the counterpart of the 
solution \(u \sim -\sqrt Q \) briefly discussed in sec.4.2.  This situation 
can in principle be systematically improved by adopting an appropriate 
asymptotic method for these solutions, which would allow us to incorporate 
them in to the shock train class simply as limiting solutions with the 
vanishing shock amplitude.

The fixed points of the map \(F(x)\) produce a discrete set of $2\pi 
$-periodic solutions.  
Each solution contains an integral number of solitons that gradually 
decrease in amplitude filling up the 
region between the shock front and the turning point.  Why does the system 
follow the solution with \(x_* = 0.497\)? It is clear that the system 
choose a steady state to follow according to the constraints imposed by the 
initial conditions.  Such a constraint is the mass invariant
 \begin{equation}
 	\bar u = \int_0^1 u(x,t=0) dx,
 	\label{caz}
 \end{equation}
which was set to zero.  In order to estimate \(\bar u \) from the lowest 
order solution (\ref{u:0}), we can to the good approximation average the 
fast oscillations and regard \(k\) as a small value for \(x > x_*\).  The 
solution \(u_0 \left (\xi(x_*,x),x \right ) \) will then produce \(\bar u = 
0 \) only for \(x_* = 1/2 \).  For arbitrary \(\bar u\)-value, the system 
follows the solution with an appropriate \(x_*\)-value.  It is interesting 
to note that a solution taken from the set \(u_0 \left (\xi (x_n,x),x 
\right ) \), being discrete fulfills the condition \(\bar u = 0 \) only 
approximately, although with a good accuracy \(\sim \lambda \ll 1 \) (here 
$x_n $ is one of the fixed points of $F(x) $).  Thus, the difference might 
in principle be compensated by the higher order terms.  We do not consider 
this problem here.  More detailed comparison of our analytical solutions 
with the numerical ones including the doublings of the spatial period will 
be done in a future paper.
\section{Discussion}
An interesting feature of the repellor considered in sec.5.2 consists in 
the concentration of its natural measure inside narrow intervals where 
\(F^\prime (x) \gg 1\).  The function \(F(x) \) maps these intervals onto 
the appreciable part of \([0,1] \) and, therefore, those are the points 
from these intervals which have better chances to avoid the escape set 
\({J_{\rm k}}\) after a number of iterates.  All the unstable fixed points, 
shown in Fig.12b belong to these intervals.  Motion on the repellor is 
essentially a random walk between these narrow intervals.  If we neglect 
their width together with the underlying multifractal structure of these 
intervals, we obtain a discretization of the shock coordinates in the 
shock-train.  The meaning of such a discretization is obvious; it 
corresponds to an integral number of vibrations or solitons between the 
shock coordinate and the turning point \(x_0\).  These regions of 
oscillatory behaviour can also be considered as optically transparent 
domains, created nonlinearly by the driver (``potential wells'').  This can 
be easily understood from the consideration of the solution near the 
turning point \(x_0 \), where the small amplitude oscillations allow the 
linear treatment (sec.3).  Figs.12b,c in particular demonstrate that at 
each second fixed point of \(F(x) \) the number of solitons differs by 
unity.  Note, that the majority of the fixed points are close to zeroes of 
function \(f(x)\) in Eq.(\ref{f=g}).  This fact suggests a simplified 
``quantization rule'' for the discrete shock coordinate \(x_* \) in the 
shock-train (which is valid for sufficiently small $g(x) $)
\begin{equation}
\int_{x_*}^{x_0}\frac {Q_0^{1/4}dx}{{\bf K} (x)(k^2+k^{\prime 
	4})^{1/4}} = \sqrt 2 \lambda \left (n + \frac{1}{4} \right )	
	\label{quant}
\end{equation}
In general, even when the parameter \(\beta \) is not large and the linear 
approximation at the turning point \(x_0\) is not justified, the number of 
solitons is a constitutive aspect of the considered chaos scenario.  To put 
it anotherway, consider the motion on the invariant set of the map 
\(F(x)\).  Suppose we know the coordinate \(x_*\) of some reference shock 
in the shock-train or, equivalently, the coordinate of the leftmost 
soliton in the correspondent potential well.  An appropriate accuracy to 
which $x_* $ should be known is \(o(\lambda ) \), as Eq.(\ref{quant}) 
suggests.  In practice, that would be enough only for identifying \(x_*\) 
with one of the region of the concentration of the natural measure, marked 
by the digits in Fig.9a.  Better accuracy in shock coordinate would allow 
us to go only few steps deeper into coarsegraining of this region.  Since 
the Lyapunov exponents are typically of order of \(1/\lambda \) on the 
repellor, we cannot even predict into which region falls the coordinate of 
the next shock in the shock-train.  With a more accurate knowledge of the 
reference shock coordinate we could at best predict a few next.  Therefore 
we will obtain an apparently random sequence of soliton numbers in the 
subsequent potential wells, created by fully deterministic driver.  In 
general, one can argue in terms of the topological conjugacy between these 
random sequences and the set of real numbers, as it is widely used in study 
of deterministic chaos in quadratic or binary tent maps, see e.g.  
\cite{dev,mcC}.  The number of the coarsegrained intervals in Fig.9a or, 
equivalently, the number of solutions of Eq.(\ref{quant}) for $x_* $ can be 
taken as a base of arithmetic.
\section{Conclusions}
Using asymptotic methods we have represented the continuous travelling wave 
solutions of the periodically driven KdV-Burgers equation as a Poincar\'e 
map at the driver period.  Unlike many other analytical approaches to 
driven equations, the amplitude of the driver is not assumed to be small.  
We argue that the typical response of this system to the external force 
consists of a fragmentation of the solution into two alternating parts.  
One of them is smooth and has the driver scale.  The second is an 
oscillatory one (shock), its scale is defined by the small dispersion and 
viscosity.  The scenario of the spatial chaotization of this solution is 
suggested.  It consists in a randomization of the shock-coordinate with 
respect to the driver phase under the action of the derived Poincar\'e map.

{\bf Acknowledgements.} The author wish to express special thanks to T.~ 
Hada, A.I.~Neishtadt, C.F.~Kennel and C.C.~Wu for interesting discussions.  
Fruitful discussions with H.D.I.~Abarbanel and the other participants of a 
seminar at the Institute for Nonlinear Sciences, La Jolla, as well as with 
F.H.~Busse and the participants of his seminar at the Bayreuth university 
are gratefully acknowledged.  This work was supported by National Science 
Foundation Grant No.~PHY~91-20591 and by the International Project ``Wave 
Chaos" sponsored by the Government of the Russian Federation.

\end{document}